\documentclass[pdflatex,sn-mathphys-num]{sn-jnl}

\usepackage{graphicx}%
\usepackage{multirow}%
\usepackage{amsmath,amssymb,amsfonts}%
\usepackage{amsthm}%
\usepackage{mathrsfs}%
\usepackage[title]{appendix}%
\usepackage{xcolor}%
\usepackage{textcomp}%
\usepackage{manyfoot}%
\usepackage{booktabs}%
\usepackage{algorithm}%
\usepackage{algorithmicx}%
\usepackage{algpseudocode}%
\usepackage{listings}%
\newgeometry{top=1in, bottom=1in, left=1.in, right=1.in}

\begin{document}

\title[Article Title]{Network structural change point detection and reconstruction for balanced neuronal networks}


\author[1,2,3,4]{\fnm{Kai} \sur{Chen}}

\author[1,2,3,4]{\fnm{Mingzhang} \sur{Wang}}

\author*[1,2,3,4]{\fnm{Songting} \sur{Li}}\email{songting@sjtu.edu.cn}

\author*[1,2,3,4]{\fnm{Douglas} \sur{Zhou}}\email{zdz@sjtu.edu.cn}

\affil[1]{\orgdiv{School of Mathematical Sciences}, \orgname{Shanghai Jiao Tong University}, \orgaddress{\city{Shanghai}, \postcode{200240}, \country{China}}}

\affil[2]{\orgdiv{Institute of Natural Sciences}, \orgname{Shanghai Jiao Tong University}, \orgaddress{\city{Shanghai}, \postcode{200240}, \country{China}}}

\affil[3]{\orgdiv{Ministry of Education Key Laboratory of Scientific and Engineering Computing}, \orgname{Shanghai Jiao Tong University}, \orgaddress{\city{Shanghai}, \postcode{200240}, \country{China}}}

\affil[4]{\orgdiv{Shanghai Frontier Science Center of Modern Analysis}, \orgname{Shanghai Jiao Tong University},
\orgaddress{\city{Shanghai}, \postcode{200240}, \country{China}}}


\abstract{
Understanding brain dynamics and functions critically depends on knowledge of the network connectivity among neurons.
However, the complexity of brain structural connectivity, coupled with continuous modifications driven by synaptic plasticity, makes its direct experimental measurement particularly challenging.
Conventional connectivity inference methods based on neuronal recordings often assumes
a static underlying structural connectivity and requires
stable statistical features of neural activities, making them unsuitable for reconstructing
structural connectivity that undergoes changes.
To fulfill the needs of reconstructing networks undergoing potential structural changes, 
we propose a unified network reconstruction framework that combines connectivity-induced change point 
detection (CPD) with pairwise time-delayed correlation coefficient (TDCC) method. 
For general neuronal
networks in balanced regimes, we develop a theoretical analysis for discriminating changes in
structural connectivity based on the fluctuation of neuronal voltage time series. We then
demonstrate a pairwise TDCC method to reconstruct the network using spike train recordings 
segmented at the detected change points. We show the effectiveness of our 
CPD-TDCC network reconstruction using large-scale
network simulations with multiple neuronal models.
Crucially, our method accommodates networks with changes in both network topologies 
and synaptic coupling strengths while retaining accuracy even with sparsely sampled subnetwork data,
achieving a critical advancement for practical applications in real experimental situations.
Our CPD-TDCC framework addresses the critical gap in network reconstruction by accounting
connectivity-induced changes points, potentially offering a valuable tool for studying
structure and dynamics in the cortical brain.
}




\maketitle

\section{Introduction}\label{sec:intro}

In neuroscience, characterizing the structural connectivity of cortical neuronal networks is
essential for better understanding the network dynamics and functions
\cite{bullmore2009complex, suarez2020linking}. However,
direct measurement of synaptic connections of cortical network in vivo remains challenging
in mammalian brains. Current experimental approaches have significant limitations:
experimentalist can access the detailed neuronal morphology and synaptic connectivity features in
small volumes of brain tissue using electronic microscopic imaging and 3D image reconstruction
algorithms \cite{themicronsconsortium2021functional},
or visualize the long-range axonal projections of a few neurons using virus tracing
techniques \cite{xu2021highthroughput}. Unfortunately, these methods are invasive and incompatible with studies in living animals.
Furthermore, synaptic connectivity in cortical networks is frequently modulated to support flexible
cognitive functions, such as learning and memory.
Synaptic plasticity, such as spike timing-dependent plasticity, can presumably alter connection
strengths or even connectivity topologies over short timescales \cite{bi1998synaptic,caporale2008spike,caroni2012structural}.
Thus, identifying such connectivity-induced change points through analyzing high dimensional activity
time series while accurately 
measuring network structural connectivity across different time periods remains an  
unresolved challenge in the field.

Meanwhile, the collective dynamical features of cortical neurons are largely influenced 
by their underlying structural connectivity. 
Taking the technological advances of large-scale recordings neuronal activity, 
the simultaneous activity recordings, such as spike trains and calcium imaging traces,
of large neuron populations is accessible. Therefore, solving such an inverse problem of using
activity data to reconstruct the underlying connectivity structure is a key research topic
in this field.
Series previous studies have developed statistical methods, such as
correlation coefficient \cite{bedenbaugh1997multiunit}, 
mutual information \cite{steuer2002mutual,li2018causal}, 
Granger causality \cite{granger1969investigating,ding2006granger,zhou2013causal}, 
and transfer entropy \cite{runge2012quantifying, schreiber2000measuring},
to measure the effective 
connectivity, seeking possibilities to infer the underlying structural connectivity.
Although effective connectivity is activity and method dependent, which is not necessarily related to
structural connectivity, previous studies reveal relationship between these two types of connectivity
under certain conditions \cite{barnett2009granger,zhou2013causal,tian2024causal}, developing 
a series of network reconstruction frameworks.
However, accurate reconstructions relying on these statistical measures only fit networks with static structural 
connectivity and require stationary activity recording, which cannot handle the system with
connectivity-induced change points.

On the other hand, the concept of change point detection (CPD), originally detecting the shifts in 
means or variances of independent samples \cite{page1954continuous}, has been widely used in many areas,
such as ecosystems \cite{hemraj2025ecosystembased}, finance \cite{pepelyshev2015realtime} and social networks \cite{kendrick2018change}. 
In neuroscience, CPD-related studies have focused for years on elucidating the temporal structure of
neural data, analyzing both spike train recordings \cite{ratnam2003changepoint,koepcke2016single,urzay2023detecting}
and large-scale neural imaging datasets \cite{marin-llobet2025neural,ondrus2025factorized},
using methods including statistical tests, graph-based modelings, and deep learning techniques
\cite{aminikhanghahi2017survey,mosqueiro2016nonparametric,zhang2021graphbased,wang2021identifying,xu2025changepoint}.
However, existing studies predominantly focus on identifying state changes manifested in time series data,
which is typically induced by switches in external sensory inputs or internal cognitive states rather than 
detecting dynamical changes caused by modulations in underlying structural connectivity. The latter present
a particularly challenging problem as connectivity-induced changes may be subtle compared to the more
pronounced state transitions typically targeted by conventional CPD methods.

In this work, we introduce a two-step network reconstruction framework for general balanced neuronal networks -- initiating
with detecting connectivity-induced change points and followed by computing pairwise time
delayed correlation coefficient (CPD-TDCC) --
addressing the challenges in reconstructing networks with change points of structural connectivity. 
First, inspired by the scaling property of balanced networks, we develop a theoretical analysis characterizing 
the fluctuations of neuronal membrane potentials, and derive a corresponding metric for 
discriminating changes of structural connectivity.
Second, after segmenting neuronal recordings with respect to detected change points,
we implement network reconstruction using time-delayed correlation coefficient (TDCC)
based on spike train recordings.
Using multiple types of neuronal network simulations as test examples, we demonstrate the effectiveness of
CPD-TDCC framework on reconstructing variations of structural connectivity matrix, capturing both
alternations in adjacency patterns and modulations in coupling weights.
CPD-TDCC-based reconstructions show significant improvements in reconstructing performances compared with
vanilla TDCC-based reconstructions for balanced-state networks with connectivity-induced change points.
Overall, our work provides novel insights into detecting subtle connectivity-induced dynamical transitions
and accurately reconstructing time-varying neuronal networks, advancing our ability to analyze
structural plasticity from observable neural activity.

\section{Methods}\label{sec:method}

\subsection{Excitatory-inhibitory balanced neuronal networks}

The balance between excitation and inhibition is ubiquitous properties
of cortical networks. The balanced neuronal network framework is first 
introduced to understand the irregular network firing 
activity characteristic across different different cortical states \cite{vanvreeswijk1996chaos,vreeswijk1998chaotic,renart2010asynchronous},
and later has been verified in electrophysiological recordings \cite{shadlen1998variable,haider2006neocortical}.
In balanced regime, the average excitatory and inhibitory inputs almost
cancel out with the average inputs remains below threshold while fluctuations drive neurons to fire spikes.
This balanced state is crucial for cortical information encoding and processing \cite{tian2020excitationinhibition,barzon2025excitationinhibition}, 
making it an essential framework for studying realistic neural network dynamics and connectivity inference methods.
Therefore, this study primarily focuses on studying spiking neuronal networks 
operating in balanced regimes.

\subsubsection{Leaky integrate-and-fire neuronal network}
We consider a network of $N$ LIF neurons, comprising $N_E$ excitatory and $N_I$ inhibitory neurons.
For $i$-th neuron with type $Q_i$, the dynamics of its membrane potential (voltage) $v_i^{Q_i}$
is defined by \cite{vanvreeswijk1996chaos}
\begin{equation}
    \begin{aligned}
        C_m\frac{\mathrm{d}v_i^{Q_i}}{\mathrm{d}t}&=-G_\mathrm{leak}(v_i^{Q_i}-v_\mathrm{leak})
        +I_{i}^{{Q_i}E}+I_i^{{Q_i}I},\\
        v_i &= v_\mathrm{reset}\quad\mathrm{if}\quad v_i(t)\ge v_\mathrm{th}^{Q_i},\\
    \end{aligned}
    \label{eq:LIF_dyn}
\end{equation} 
where $C_m$ represents the neuron's membrane capacitance, $G_\mathrm{leak}$ and
$v_\mathrm{leak}$ are the leaky conductance and leaky reversal potential, respectively.
${Q_i}\in \{E,I\}$ denotes the neuron type (excitatory or inhibitory).
$v^{Q_i}_\mathrm{th}$ is the spike threshold for type $Q_i$ neurons.
$I_i^{Q_iE}$ and $I_i^{Q_i I}$ represent the total excitatory and inhibitory synaptic currents received by neuron $i$, respectively. 
Upon reaching its specific spike threshold $v^{Q_i}_\mathrm{th}$, $i$-th neuron generates an
action potential (spike) that is transmitted to all its post-synaptic neurons.
Immediately following that, its voltage is reset to $v_\mathrm{reset}$. 
To simplify the analysis, we set network parameters as dimensionless values,
retaining only the dimension of time:
$C_m=1$, $v_\mathrm{reset}=0$, $v^E_\mathrm{th}=1$, $v^I_\mathrm{th}=0.7$, $G_\mathrm{leak}=0.05\,\textrm{ms}^{-1}$.

The excitatory input current to the $i$-th neuron with type $Q_i$ is given by
\begin{equation}
I_i^{Q_iE}(t)=F_i^{Q_i} \sum_k \delta\left(t-s_{i,k}^{Q_i}\right)+\sum_{j=1}^{N_E} W_{ij}^{{Q_i}E} \sum_k \delta\left(t-\tau_{j,k}^E\right),  
\label{eq:E_current}
\end{equation}
where $\delta(\cdot)$ is the Dirac delta function.
The first term represents the external homogeneous Poisson input to the $i$-th neuron
where $s_{i,k}$ is the $k$-th spike time, $F_i$ is the input strength, and $\nu_i$ is the input
rate. The second term describes the recurrent excitatory inputs from within the network, where 
$\tau^E_{j,k}$ is the $k$-th spike time of the $j$-th excitatory neuron, and $W_{ij}^{Q_iE}$ 
is the coupling strength from $j$-th excitatory neuron to $i$-th neuron.
The inhibitory input current to the $i$-th neuron with $Q_i$ arises from recurrent interactions 
with inhibitory neurons within the network, and is defined by 
\begin{equation}
I_i^{Q_i I}=\sum_{j=1}^{N_I} W_{i j}^{Q_i I} \sum_k \delta\left(t-\tau_{j,k}^I\right),
\label{eq:I_current}
\end{equation}
where $\tau^I_{j,k}$ is the $k$-th spike time of the $j$-th inhibitory neuron, and
$W_{ij}^{Q_iI}$ is the coupling strength from the $j$-th inhibitory neuron to the $i$-th neuron.

To investigate the network dynamics in the excitatory-inhibitory (EI) balanced
regime, we adopt the parameter settings for coupling weights and Poisson inputs from previous 
works \cite{gtkc18dynamics}.
In this setup, each neuron receives $K$ excitatory and $K$ inhibitory connections, initialized
randomly, on average from other neurons in the balanced network.
The recurrent coupling strength from neurons of type $Q_i$ to those of type $Q_j$, denoted
by $W_{ij}^{Q_iQ_j}$, is scaled by the degree parameter $K$ according to the form
$W_{ij}^{Q_iQ_j}=J^{Q_iQ_j}/\sqrt{K}$, a scaling often used to ensure finite fluctuations in the 
large $K$ limit.
Similarly, the strength of the Poisson input to the $i$-th neuron, denoted by $F^{Q_i}$, is 
scaled as $F^{Q_i}=J^{Q_i}/\sqrt{K}$, with the Poisson input rate given by $\nu^{Q_i}=\mu_0K$.
To satisfy the balanced condition, which requires a delicate cancellation between excitatory and
inhibitory inputs, the coupling strength are chosen such that
$$\frac{J^I}{J^E}>\frac{J^{II}}{J^{EI}}>\frac{J^{IE}}{J^{EE}}.$$
The detailed choice of $J^{Q_iQ_j}$, $J^{Q_i}$ and $\nu_0$ will be introduced later in the result section.


In this work, we numerically simulate all networks using exponential Euler scheme with a small
time step of $\Delta t=0.02$ ms. These simulations were efficiently implemented using
BrainPy \cite{wang2023brainpy}, a GPU-accelerated simulator for neuronal 
dynamics, which allows for the efficient simulation of large-scale spiking neuronal networks.
The voltage time series and spike trains of a 200-neuron randomly sampled subnetwork for each network are recorded for further reconstruction analysis.

\subsubsection{Hodgkin-Huxley neuronal network}
The dynamics of the $i$-th neuron of a Hodgkin-Huxley (HH) network is defined by \cite{hodgkin1952quantitative,sun2009librarybased}
\begin{equation}\label{eq: V of HH}
C_m\frac{dv_{i}^{Q_i}}{dt}=-G_{l}(v^{Q_i}_{i}-v_{l})+I^\textrm{Na}_{i}+I^\textrm{K}_{i}+I_{i}^{{Q_i}E} + I_i^{{Q_i}I},
\end{equation}
where $C_m$ is the neuron's membrane capacitance, and $v_i^{Q_i}$ is the voltage of
$i$-th neuron with type ${Q_i}$ (${Q_i}\in\{E,I\}$).
Similar to LIF model, $G_l$ is the conductance of leaky current and $v_l$ 
is the corresponding reversal potential. $I_{i}^{{Q_i}E}$ and $I_i^{{Q_i}I}$ are the excitatory and inhibitory input
currents, respectively, with the same form as defined by Eqs. \ref{eq:E_current} and \ref{eq:I_current}.
$I_i^\mathrm{Na}$ and $I_i^\mathrm{K}$ are the sodium and potassium ionic currents, defined by
\begin{equation}
\begin{aligned}
I_i^\textrm{Na} &= -G_\textrm{Na}m_{i}^{3}h_{i}(v_{i}-v_\textrm{Na}),\\
I_i^\textrm{K} &= -G_\textrm{K}n_{i}^{4}(v_{i}-v_\textrm{K}),\\
\frac{dz_{i}}{dt}&=(1-z_{i})\alpha_{z}(v_{i})-z_{i}\beta_{z}(v_{i}),\,\,\,\text{ for }z=m,h,n.
\end{aligned}
\end{equation}
Here, the cell-type superscript of $v_i$ is dropped for simplicity, since the expressions
ionic current are identical for both excitatory and inhibitory neurons.
Similar to the leaky current, 
$G_\textrm{Na}$ and $G_\textrm{K}$ are the maximum conductance for sodium and potassium current, and
$v_\textrm{Na}$ and $v_\textrm{K}$ are the corresponding reversal potentials, respectively.
$m_{i}$, $h_{i}$, and $n_{i}$ are gating variables for conductances, with the voltage-dependent
rate variable $\alpha_{z}$ and $\beta_{z}$, defined by
\cite{dayan2005theoretical}
\begin{equation}
\begin{aligned}\alpha_{m}(v) & =\frac{0.1v+4}{1-\exp(-0.1v-4)}, &  & \beta_{m}(v)=4\exp\left(\frac{-(v+65)}{18}\right),\\
\alpha_{h}(v) & =0.07\exp\left(\frac{-(v+65)}{20}\right), &  & \beta_{h}(v)=\frac{1}{1+\exp(-3.5-0.1v)},\\
\alpha_{n}(v) & =\frac{0.01v+0.55}{1-\exp(-0.1v-5.5)}, &  & \beta_{n}(v)=0.125\exp\left(\frac{-(v+65)}{80}\right).
\end{aligned}
\end{equation}
When the voltage $v_{i}$ reaches the spiking threshold $v_\mathrm{th}$ at $t=\tau_{i,k}$, the $i$-th neuron 
fires its $k$-th spike, and sends pulse inputs to its connected neurons following Eqs. \ref{eq:E_current} and
\ref{eq:I_current}.
The network parameters are defined as follows:
$C_m=1\,\mu\mathrm{F}$,
$v_\mathrm{Na}=50$ mV, $v_\mathrm{K}=-77$ mV, $v_{L}=-54.387$ mV, $v_\mathrm{th}=20\,\mathrm{mV}$,
$G_\mathrm{Na}=120\,\mathrm{mS}\cdot\mathrm{cm}^{-2}$,
$G_\mathrm{K}=36\,\mathrm{mS}\cdot\mathrm{cm}^{-2}$ and
$G_{L}=0.3\,\mathrm{mS}\cdot\mathrm{cm}^{-2}$. 

Parameter scalings for coupling strengths and Poisson inputs to fulfill balanced-state conditions are defined
similarly as those in LIF networks, and the detailed values of $J^{Q_iQ_j}$, $J^{Q_i}$ and $\nu_0$ will be
covered in the result section.


\subsubsection{Morris-Lecar neuronal network}
The dynamics of the $i$-th neuron with type ${Q_i}$ (${Q_i}\in\{E,I\}$) in a Morris-Lecar network is governed by \cite{morris1981voltage}

\begin{equation*}
    \begin{aligned}
        C_m \frac{d v^{Q_i}_i}{d t} =&-G_\mathrm{Ca} M^{\infty}_i\left(v_i^{Q_i}-v_\mathrm{Ca}\right) -G_{K} W_i\left(v^{Q_i}_i-v_\mathrm{K}\right)\\
        &-G_{l}\left(v^{Q_i}_i-v_{l}\right)+I^\mathrm{{Q_i}E}_i + I^\mathrm{{Q_i}I}_i \\
        \frac{d W_i}{d t} =&\frac{1}{\tau^W_i}\left(W_i^{\infty}-W_i\right),
    \end{aligned}
\end{equation*}
where $C_m$ is the neuron's membrane capacitance, and $v_{i}^{Q_i}$ is neuron's membrane potential.
$v_\mathrm{Ca},v_\mathrm{K}$, and $v_{l}$ are the reversal potentials for the calcium, potassium, and leak currents, respectively. $G_\mathrm{Ca},G_\mathrm{K}$ and $G_{l}$ are the corresponding maximum conductance.
$W_i$ is the neuron's recovery variable (normalized potassium conductance).
$M^\infty_i$ and $W^\infty_i$ are the voltage-dependent equilibrium value of the normalized calcium and potassium conductance, respectively, defined by
\begin{equation*}
    \begin{aligned}
        M_i^{\infty}=&0.5\left(1+\tanh{[(v^{Q_i}_i-v_1)/v_2]}\right) \\
        W_i^{\infty}=&0.5\left(1+\tanh{[(v^{Q_i}_i-v_3)/v_4]}\right) \\
    \end{aligned}
\end{equation*}
where $v_1$, $v_2$, $v_3$, and $v_4$ are the model parameters. $\tau_i^W$ is the
voltage-dependent time constant of $W_i$, defined by
\begin{equation*}
    \tau^W_i=\tau_0\left(\cosh{\frac{v^{Q_i}_i-v_3}{2v_4}}\right)^{-1},
\end{equation*}
where $\tau_0$ is a temperature-dependent parameter, regarding as a constant in our simulations.
$I_i^{Q_iE}$ and $I_i^{Q_iI}$ are excitatory and inhibitory input currents to $i$-th neuron,
defined the same as those in the LIF model by Eqs. \ref{eq:E_current} and \ref{eq:I_current}.
When $v_i^{Q_i}$ reaches threshold $v_\mathrm{th}$, $i$-th neuron emits a spike to all its post-synaptic neurons.
In numerical simulation, the network parameters are chosen as: 
$G_\mathrm{Ca}=4.4\,\mathrm{mS\cdot cm^{-2}}$,
$G_\mathrm{K}=8\,\mathrm{mS\cdot cm^{-2}}$,
$G_{l}=2\,\mathrm{mS\cdot cm^{-2}}$,
$v_\mathrm{Ca}=130\,\mathrm{mV}$, 
$v_\mathrm{K}=-84\,\mathrm{mV}$,
$v_{l}=-60\,\mathrm{mV}$,
$C=20\,\mathrm{mF\cdot cm^{-2}}$,
$v_1=-1.2$ mV, $v_2=18$ mV, $v_3=2$ mV, $v_4=30$ mV,
$\tau_0=40$ ms and $v_\mathrm{th}=10$ mV.

Parameter scalings for coupling strengths and Poisson inputs to Morris-Lecar networks
are defined by the same expressions as those in LIF networks, and the detailed values of
$J^{Q_iQ_j}$, $J^{Q_i}$ and $\nu_0$ will be
covered in the result section.


\subsection{Network reconstruction with pulse-output signals}\label{sec:recon_theory}

Pearson correlation coefficient (CC) is one of the most commonly used tool for
measuring the functional correlation between signals \cite{bedenbaugh1997multiunit}.
However, due to the symmetric nature of CC, it cannot capture the directional causal
relation between signals, while only captures the simultaneous correlation between signals.
To address this, artificial time delays are introduced \cite{ito2011extending, tian2024causal,laasch2025comparison}.
Consider a pair of neurons, denoted as $X$ and $Y$, whose activities are measured as time
series $\{x_t\}$ and $\{y_t\}$. Assuming those time series are wide-sense stationary \cite{florescu2014probability}, 
the time delayed correlation coefficient (TDCC) from $X$ to $Y$ is defined by
\begin{equation}
    \mathrm{TDCC}_{X\to Y}(\tau) = \frac{\mathbb{E}[(x_{t-\tau}-\mathbb{E}(x_{t-\tau}))(y_t-\mathbb{E}(y_t))]}
    {\sqrt{\mathbb{E}\left[(x_{t-\tau}-\mathbb{E}(x_{t-\tau}))^2\right]\mathbb{E}\left[(y_t-\mathbb{E}(y_t))^2\right]}},
    \label{eq:tdcc}
\end{equation}
where $\mathbb{E}(\cdot)$ denotes the expectation of a random variable.
TDCC with a positive (negative) $\tau$ indicates the correlation between the past (future)
state of $X$ and the present state of $Y$, measuring the causal effect from $X$ to $Y$ 
($Y$ to $X$). And TDCC with a zero-$\tau$ reduces to classical CC.

In this work, we apply TDCC directly onto the spike-train time series instead of voltage
time series. Our previous works have shown that the TDCC values are quantitatively related to other more complex causality measures, such as Granger causality and transfer entropy, when applying to binary valued spike-train time series
of neuronal networks \cite{tian2024causal}.
Therefore, taking the merit of simplicity and effectiveness of TDCC, we 
compute pairwise TDCC cross all neuronal pairs in large neuronal
network, and use TDCC values as the metrics for network reconstruction.
In this work, we mainly focused on reconstructing the adjacency matrix of 
networks, \textit{i.e.}, the existence of connection or not given a 
neuronal pair.
For such a binary classification problem,
the reconstruction performance is measured using the receiver operating 
characteristic (ROC) curves. The area under the curve (AUC) quantifies the 
performance of the reconstruction. If the AUC is close to $1$, indicating
a well separation between metric distribution of connected and unconnected
neuronal pairs, the reconstruction accuracy is almost 100\%.
Otherwise, if the AUC is around $0.5$, indicating the metric distribution of 
connected and unconnected pairs are indistinguishable, the reconstruction
performance is almost the same as random guess.

Here, we also derive the quantitative relationship between the TDCC value and 
the coupling strength of synaptic interactions for the binary valued
spike-train time series.
For two binary valued spike-train time series $\{x_t\}$ and $\{y_t\}$, 
the causal relation from $X$ to $Y$ can be interpreted as the change of 
firing probability ($p(y_t=1)$) induced by the $X$'s spike ($x_{t-\tau}=1$),
defining as
\begin{equation}
    \Delta p_{X\to Y}(\tau) = p(y_t=1|x_{t-\tau}=1) - p(y_t=1|x_{t-\tau}=0).
    \label{eq:causality}
\end{equation}
This interpretation of causality between two spike-train time series
is consistent with the definition of causality from Judi Pearl \cite{pearl1994probabilistic}.
Generally, $\Delta p_{X\to Y}$ is function of the coupling 
strength from $X$ to $Y$, \textit{i.e.}, $\Delta p_{X\to Y}=f(S_{X\to Y})$.
Taking Taylor expansion with respect to $S_{X\to Y}$, we have
\begin{equation}
    \Delta p_{X\to Y} = f(0) + \left.\frac{\mathrm{d}f}{\mathrm{d}S}\right|_{S_{X\to Y}} S_{X\to Y} + \mathcal{O}(S_{X\to Y}^2).
    \label{eq:dp_taylor}
\end{equation}
If $S_{X\to Y}=0$, then neuron $X$ and $Y$ are causaly independent with each
other, and $\Delta p_{X\to Y}=0$, leading to $f(0)=0$.
Therefore, considering the weak coupling regime, \textit{i.e.}, $|S_{X\to Y}| \ll 1$, $\Delta p_{X\to Y}$ is linearly proportional to synaptic 
coupling strength.

For binary valued time series, we can expend Eq. \ref{eq:tdcc} with
respect to all possible states in joint probability space as
\begin{equation}
\mathrm{TDCC}_{X\to Y}(\tau) = \frac{p(x_{t-\tau}=1, y_t=1)-p(x_{t-\tau}=1)
p(y_t=1)}{\sqrt{p(x_{t-\tau}=1)p(x_{t-\tau}=0)}\sqrt{p(y_t=1)p(y_t=0)}}.
\label{eq:tdcc_dp}
\end{equation}

To match the expression of TDCC, we rewrite the Eq. \ref{eq:causality} as
follows
\begin{equation}
\begin{aligned}
    \Delta p_{X\to Y}(\tau) &= 
    \frac{p(y_t=1,x_{t-\tau}=1)}{p(x_{t-\tau}=1)} -
    \frac{p(y_t=1,x_{t-\tau}=0)}{p(x_{t-\tau}=0)}\\
    &= \frac{p(y_t=1,x_{t-\tau}=1)(1-p(x_{t-\tau}=1))-p(y_t=1,x_{t-\tau}=0)p(x_{t-\tau}=1)}{p(x_{t-\tau}=1)p(x_{t-\tau}=0)}\\
    &= \frac{p(y_t=1,x_{t-\tau}=1)-p(y_t=1)p(x_{t-\tau}=1)}{p(x_{t-\tau}=1)p(x_{t-\tau}=0)}.
\end{aligned}
\label{eq:dp_derivation}
\end{equation}
Comparing Eqs. \ref{eq:tdcc_dp} and \ref{eq:dp_derivation}, we can show that
$\Delta p_{X\to Y}(\tau)$ is linearly related to $\mathrm{TDCC}_{X\to Y}(\tau)$:
\begin{equation}
\mathrm{TDCC}_{X\to Y}(\tau) = \Delta p_{X\to Y}(\tau)\cdot \sqrt{\frac{p(x_{t-\tau}=1)p(x_{t-\tau}=0)}{p(y_t=1)p(y_t=0)}}
\label{eq:linear_relation}
\end{equation}
Here, we denote the mean firing rate of $X$ and $Y$ as $r_x$ and $r_y$, respectively. Thus, $p(x_t=1)=r_x\Delta t$ and $p(y_t=1)=r_y\Delta t$, where
$\Delta t$ is the time step used for converting spike trains to binary time
time series data. According to the wide-sense stationary assumption, the
neuronal mean firing rate is invariant across time, and 
$p(x_{t-\tau})=p(x_\tau)$ for binary variable $\{x_t\}$.
Thus, we can further simplify the Eq. \ref{eq:linear_relation} as
\begin{equation}
\mathrm{TDCC}_{X\to Y}(\tau) = \Delta p_{X\to Y}(\tau)\cdot 
\sqrt{\frac{r_{x}\cdot(1-r_x\Delta t)}{r_{y}\cdot(1-r_y\Delta t)}}
\label{eq:linear_relation2}
\end{equation}
Combining Eqs. \ref{eq:dp_taylor} and \ref{eq:linear_relation2}, we show that
pairwise TDCC value, estimated using binary spike-train time series,
is linearly related to underlying synaptic coupling strength in the weak
coupling regime.

\section{Results}\label{sec:result}

\subsection{Connectivity-induced change point detections in neuronal networks}

In this work, we mainly focus on investigating neuronal networks that undergo 
rapid structural changes at specific time points.
Such abrupt transitions can arise from biological mechanisms like synaptic plasticity and structural plasticity,
which are known to operate on much shorter timescales than the typical duration of neural recordings. 

We consider an $N$-neuron neuronal network with an initial connectivity matrix $\mathbf{W}_1$.
At a specific time $t=t_c$, the connectivity matrix is instantaneously switched to a new matrix $\mathbf{W}_2$,
while all other network parameters remain unchanged.
Our investigation focuses on two distinct scenarios of connectivity changes:
\begin{enumerate}
    \item Topology change with homogeneous weights: $\mathbf{W}_1$ and $\mathbf{W}_2$ share the same homogeneous (cell-type dependent) coupling 
    strengths but differ in their adjacency matrices, with both maintaining
    the same average in-degree. This scenario models structural plasticity, where synaptic connections are rewired without changing the synaptic strength distribution \cite{caroni2012structural}.
    \item Weight change with fixed network topology: $\mathbf{W}_1$ and $\mathbf{W}_2$ shares an identical adjacency matrix but differ in heterogeneous coupling strengths, both
    drawn from the same distribution. This transition models synaptic plasticity, where the connection pattern is preserved but the strengths of existing synapses are modified \cite{bi1998synaptic}.
\end{enumerate}

In both scenarios, the structural changes significantly impact the network dynamics, posing challenges for accurately reconstructing the underlying connectivity.
In the following sections, we introduce a novel change point detection (CPD) method to identify the timing of such structural transitions, followed by strategies for reconstructing the corresponding connectivity matrices before and after the change.

\subsection{Voltage fluctuations in balanced state neuronal networks}
To provide a broader context, we now consider a general spiking neuronal network where the
dynamics of $i$-th neuron, driven by homogeneous Poisson inputs with strength 
$F_i$ and rate $\nu$, can be described by
\begin{equation}
C_m \frac{d v_i}{d t}=I_{i,\mathrm{ion}}
+F_i\sum_k \delta\left(t-s_{i, k}\right)
+\sum_{j=1}^{N} W_{i j} \sum_k \delta\left(t-t_{j, k}\right).
\label{eq:general_neu_ode}
\end{equation}
The first term $I_{i,\mathrm{ion}}$ represents the total transmembrane ionic current
for neuron $i$, which corresponding to leaky current in LIF neurons, leaky, sodium,
and potassium current in HH neurons, leaky, calcium, and potassium current in
Morris-Lecar neurons.
The remaining terms are the input current from external Poisson input and recurrent interactions
within the network, respectively.
These terms directly correspond to the input currents detailed in Eqs.
\ref{eq:E_current} and \ref{eq:I_current}.

We integrate the voltage in Eq. \ref{eq:general_neu_ode} over $[t,t+\Delta t]$, 
yielding:
\begin{equation}
\begin{aligned}
C_m \left[v_i(t+\Delta t)-v_i(t)\right] =& \int_t^{t+\Delta t}
I_{i,\mathrm{ion}}(t')dt' 
+ F_i\sum_k\mathbf{1}_{[t\le t_{j,k}<t+\Delta t]}\\
&+ \sum_{j=1}^{N}W_{ij}\sum_k\mathbf{1}_{[t\le t_{j,k}<t+\Delta t]},
\end{aligned}
\label{eq:dv}
\end{equation}
where $\mathbf{1}_{[t\le t_{j,k}<t+\Delta t]}$ is the indicator function
for the spike time of $j$-th neurons, and $\mathbf{1}_{[t\le t_{i,k}<t+\Delta t]}$ is the indicator function for Poisson inputs for $i$-th 
neuron.
Given the refractoriness of biological neurons and by choosing a relatively
small time window $\Delta t$, \textit{e.g.} $0.5$ ms, there is at most one spiking
event happens within $[t, t+\Delta t]$. Therefore, we can approximate the
summation of indicator functions $\mathbf{1}_{[t\le t_{j,k}<t+\Delta t]}$
with independent Bernoulli random variables, $d_j(t)$, where the probability of a spike occurring 
is $p(d_j(t)=1)=m_j\Delta t$, with $m_j$ being the $j$-th neuron's mean firing rate.
Similarly, we approximate $\mathbf{1}_{[t\le t_{i,k}<t+\Delta t]}$ with independent
Bernoulli random variable $d^P_i(t)$, representing the occurrence of a Poisson input
to the $i$-th neuron, with probability $p(d^P_i(t)=1)=\nu_i\Delta t$, where
$\nu_i$ is the rate of homogeneous Poisson input process.

Average Eq. \ref{eq:dv} over all neurons, we have
\begin{equation}
\begin{aligned}
    \frac{C_m}{N} \sum_{i=1}^N\left[v_i(t+\Delta t)-v_i(t)\right] =& 
    \frac{1}{N} \sum_{i=1}^N\int_t^{t+\Delta t}
    I_{i,\mathrm{ion}}(t')dt' \\
    &+ \frac{1}{N} \sum_{i=1}^NF_id_i^P(t)
    + \frac{1}{N} \sum_{i=1}^N\sum_{j=1}^{N}W_{ij}d_j(t).
\end{aligned}
\label{eq:dv_int}
\end{equation}
Under the large-$N$ limit, \textit{i.e.}, $N\to\infty$,
according to central limit theorem, the distribution of the second term 
in Eq. \ref{eq:dv_int} probabilistically approaches to normal distribution 
as
\begin{equation}
    \frac{1}{N} \sum_{i=1}^NF_id_i^P(t)
    \xrightarrow{\mathcal{D}}
    \mathcal{N}\left(\frac{1}{N} \sum_{i=1}^NF_i\bar\nu\Delta t,
    \frac{1}{N^2}\sum_{i=1}^NF_i^2\bar\nu\Delta t\right).
\end{equation}
Similarly, the distribution of third term in Eq. \ref{eq:dv_int}, in large-$N$ limit,
also converges to normal distribution, yielding:
\begin{equation}
\begin{aligned}
    \frac{1}{N} \sum_{i=1}^{N}\sum_{j=1}^NW_{ij}d_j(t) &=
    \frac{1}{N} \sum_{j=1}^{N}\left(\sum_{i=1}^NW_{ij}\right)d_j(t)
    =\frac{1}{N} \sum_{j=1}^{N}\hat{W}_{*j}d_j(t)\\
    &\xrightarrow{\mathcal{D}}
    \mathcal{N}\left(\frac{1}{N} \sum_{j=1}^{N}\hat{W}_{*j}m_j\Delta t,
    \frac{1}{N^2} \sum_{j=1}^{N}\hat{W}_{*j}^2m_j \Delta t\right).
\end{aligned}
\label{eq:W_dist}
\end{equation}
Here, $m_i$ is the $i$-th neuron's mean firing rate,
and $\hat{W}_{*j}$ represents the summation of out-degree weights for the $j$-th neuron.

To centralize these terms in Eq. \ref{eq:dv_int}, we take the second-order 
differentiation of voltage, $\Delta^2v_i:=v_i(t+2\Delta t)-2v_i(t+\Delta t)+v_i(t)$,
yielding:
\begin{equation}
\begin{aligned}
    \frac{C_m}{N} \sum_{i=1}^N\Delta^2v_i 
    &= \frac{1}{N} \sum_{i=1}^N\int_t^{t+\Delta t}
    (I_{i,\mathrm{ion}}(t'+\Delta t)-I_{i,\mathrm{ion}}(t'))dt' \\
    &\quad+ \frac{1}{N} \sum_{i=1}^NF_i\left[d_i^P(t+\Delta t)-d_i^P(t)\right]
    + \frac{1}{N} \sum_{j=1}^{N}\hat{W}_{*j}\left[d_j(t+\Delta t) - d_i(t)\right].
\end{aligned}
\label{eq:ddv}
\end{equation}
Assuming the ionic current is Lipschitz continuous, we estimate the first term in 
Eq. \ref{eq:ddv} as the higher order infinitesimal of $\Delta t$, as
\begin{equation}
\begin{aligned}
|\langle\Delta^2v_i^\mathrm{ion}\rangle_i|:=&\left|\frac{1}{N} \sum_{i=1}^N \int_t^{t+\Delta t}-\left(I_{i, \text { ion }}(t'+\Delta t)-I_{i, \mathrm{ion}}(t')\right) \mathrm{d} t'\right|\\
&\leq \frac{1}{N} \sum_{i=1}^N \int_t^{t+\Delta t}\left|\left(I_{i, \mathrm{ion}}(t'+\Delta t)-I_{i, \mathrm{ion}}(t')\right)\right| \mathrm{d} t' \\
&\leq L(\Delta t)^2,
\end{aligned}
\end{equation}
where $L$ is the Lipschitz constant.
The distributions of the second and third terms are given by
\begin{equation}
\begin{aligned}
\langle\Delta^2v_i^\mathrm{ext}\rangle_i:=\frac{1}{N} \sum_{i=1}^NF_i\left[d_i^P(t+\Delta t)-d_i^P(t)\right]
&\xrightarrow{\mathcal{D}} \mathcal{N}\left(0, \frac{2}{N^2}\sum_{i=1}^NF_i^2\bar\nu\Delta t\right), \\
\langle\Delta^2v_i^\mathrm{rec}\rangle_i:=\frac{1}{N} \sum_{j=1}^{N}\hat{W}_{*j}\left[d_j(t+\Delta t) - d_i(t)\right]
&\xrightarrow{\mathcal{D}} \mathcal{N}\left(0, \frac{2}{N^2} \sum_{j=1}^{N}\hat{W}_{*j}^2m_j \Delta t \right).
\end{aligned}
\label{eq:sigma_ext_rec}
\end{equation}
According to the scaling of EI-balanced network, we have 
\begin{equation*}
\begin{aligned}
&F_i\sim\mathcal{O}\left(\frac{1}{\sqrt{K}}\right), \quad 
\bar\nu\sim\mathcal{O}\left(K\right), \\
&\hat{W}_{*j} = \sum_{i=1}^NW_{ij}\sim \mathcal{O}(K)\cdot \mathcal{O}\left(\frac{1}{\sqrt{K}}\right) = \mathcal{O}\left(\sqrt{K}\right), \quad m_j\sim\mathcal{O}(1).
\end{aligned}
\end{equation*}
The variance of the second and third terms in Eq. \ref{eq:ddv} have the order 
$\mathcal{O}\left(\sqrt{\Delta t/N}\right)$ and $\mathcal{O}\left(\sqrt{\Delta t\cdot K/N}\right)$, respectively.
Therefore, for the network with fixed connection density $p=K/N$, 
at the large-$K$ limit, the third term, contributed by the recurrent
interactions within the network, dominants the second order differentiation of voltage,
\begin{equation}
    \langle\Delta^2v_i\rangle_i \sim \mathcal{N}(0,\sigma^2_{\Delta^2v}), \qquad 
    \mathrm{with}\quad\sigma_{\Delta^2v}=\frac{1}{C_m}\sqrt{\frac{2}{N}\sum_{j=1}^N\hat{W}_{*j}^2m_j\Delta t}.
    \label{eq:sigma_rec}
\end{equation}

\begin{figure}[!h]
\centering
\includegraphics[width=\textwidth]{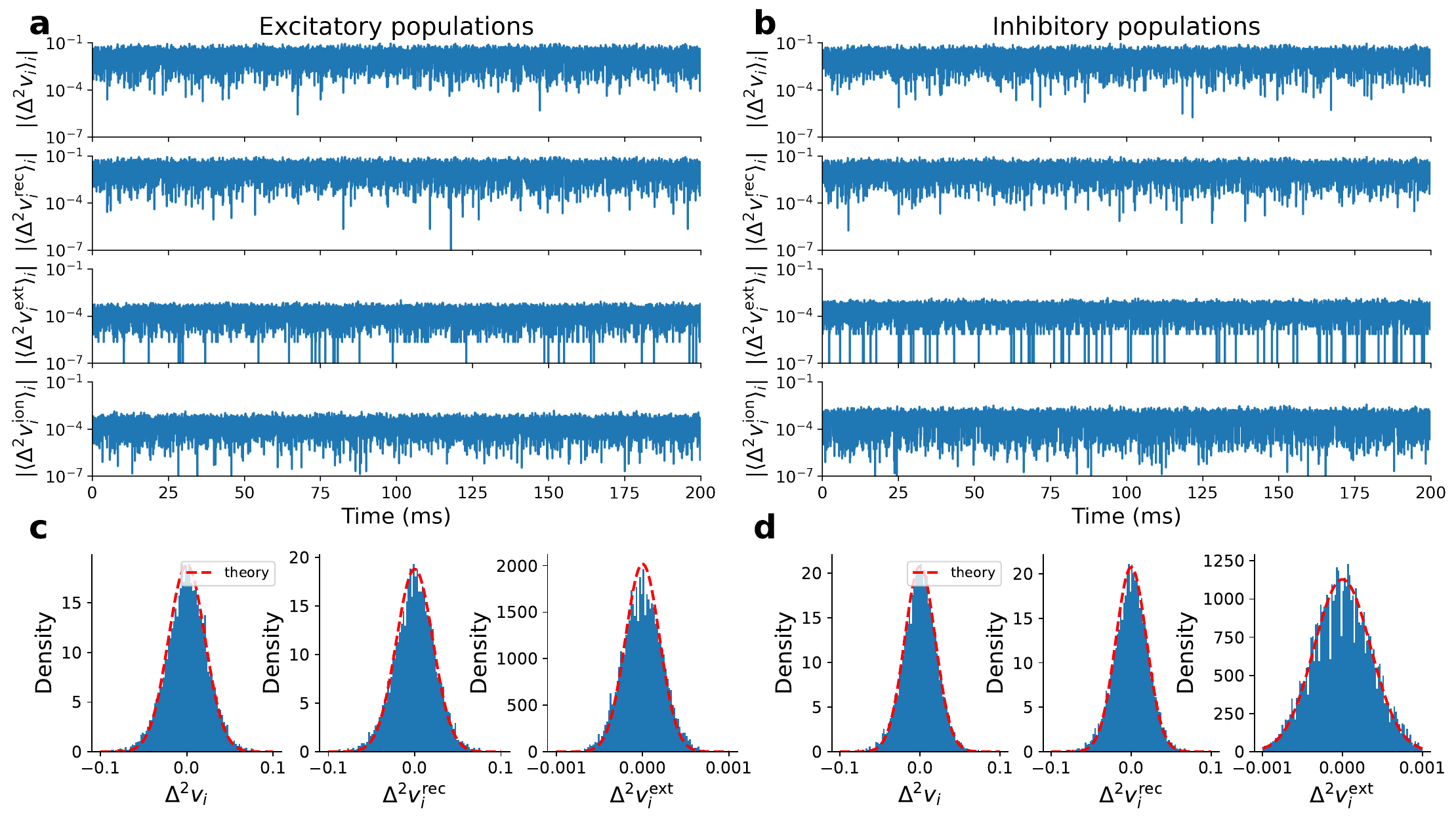}
\caption{
Population averaged second-order voltage differences
for LIF neuronal network in balanced state.
\textbf{a}. The absolute value time series of second-order voltage differences, averaged over excitatory 
populations. Subfigures (from top to bottom): raw voltage, contribution
from recurrent interactions, contribution from external Poisson inputs, 
and contribution from
intrinsic ionic current of neuronal models.
\textbf{b}. Equivalent analysis (as those in \textbf{a}) for inhibitory populations.
\textbf{c}. Histograms of population averaged second-order voltage differences,
computed from (from left to right)
raw voltage time series,
recurrent interaction contributions, and external Poisson input contributions, respectively,
for excitatory populations.
\textbf{d}. Equivalent analysis (as those in \textbf{c}) for inhibitory populations.
}
\label{fig:ddv4k}
\end{figure}

To numerically verify the variance estimation above, we simulate an LIF network 
consisting of 25600 excitatory (E) and 6400 inhibitory (I) neurons, with degree 
parameter $K=320$, and record $200$ ms voltage time series for all neurons. 
Network parameters are selected according to Table \ref{tab1}.
We compute the magnitude of $\langle\Delta^2 v_i\rangle_i$ averaged over excitatory and inhibitory neuron 
populations, respectively, as 
shown in top panels of Fig. \ref{fig:ddv4k}\textbf{a}-\textbf{b}.
We decompose the contribution of $\langle\Delta^2 v_i\rangle_i$ from recurrent interactions,
external Poisson inputs and transmembrane ionic currents separately, as shown in the 
remaining panels in Fig. \ref{fig:ddv4k}\textbf{a}-\textbf{b}, 
respectively. 
As predicted by our theoretical analysis, the variance contributed by 
$\Delta^2v_i^\mathrm{rec}$ demonstrates a predominant role, with magnitude orders
greater than contributions from the remaining terms.

The histogram of $\langle\Delta^2v_i\rangle_i$, $\langle\Delta^2v_i^\mathrm{rec}\rangle_i$ and 
$\langle\Delta^2v_i^\mathrm{ext}\rangle_i$ are presented in Fig. \ref{fig:ddv4k}\textbf{c}-\textbf{d} for excitatory and inhibitory neuron populations, respectively. 
We estimate the summation of out-degree weights $\hat{W}_{*j}$ for excitatory ($\hat{W}_{*E}$) and inhibitory ($\hat{W}_{*I}$) neurons, under the scaling of balanced 
regimes, as follows:
\begin{equation}
\begin{aligned}
    \hat{W}_{*E} &= \frac{1}{\sqrt{K}}\left(J^{EE}N_E\cdot\frac{K}{N_E}+J^{IE}N_I\cdot\frac{K}{N_E}\right)
    = \sqrt{K}\left(J^{EE}+J^{IE}\cdot\frac{N_I}{N_E}\right), \\ 
    \hat{W}_{*I} &= \frac{1}{\sqrt{K}}\left(J^{EI}N_E\cdot\frac{K}{N_I}+J^{II}N_I\cdot\frac{K}{N_I}\right) 
    = \sqrt{K}\left(J^{EI}\cdot\frac{N_E}{N_I}+J^{II}\right). \\ 
\end{aligned}
\label{eq:W_rec}
\end{equation}
By plugging in the parameters for LIF networks listed in Table \ref{tab1},
we demonstrate that our analytical predictions for the variances,
as defined in Eqs. \ref{eq:sigma_ext_rec} and \ref{eq:sigma_rec}, accurately capture
those observed in numerical simulations, as shown by the red dashed lines in Fig. \ref{fig:ddv4k}\textbf{c}-\textbf{d}.

\begin{table}[h]
\begin{tabular*}{\textwidth}{@{\extracolsep\fill}lcccc}
\toprule%
& & \multicolumn{3}{@{}c@{}}{Value} \\
\cmidrule{3-5}%
Parameter & Notation & LIF & HH & Morris-Lecar \\
\midrule
Strength factor from E to E neurons & $J^{EE}$ & 1.0   & 10 mV  & 200 mV  \\
Strength factor from E to I neurons & $J^{IE}$ & 1.0   & 10 mV  & 200 mV  \\
Strength factor from I to E neurons & $J^{EI}$ & -2.0  & -20 mV & -400 mV \\
Strength factor from I to I neurons & $J^{II}$ & -1.8  & -18 mV & -360 mV \\
Poisson Strength to E neurons & $J^{E}$ & 1.0  & 10 mV & 200 mV  \\
Poisson Strength to I neurons & $J^{I}$ & 0.8  & 8 mV  & 160 mV  \\
Rate scaling of external Poisson inputs & $\nu_0$ & 50 Hz & 50 Hz & 50 Hz \\
\botrule
\end{tabular*}
\centering
\caption{Network parameters for different types of networks in the balanced regime.}\label{tab1}
\end{table}

\subsection{Change-point detection based voltage fluctuations}\label{sec:CPD}

\begin{figure}[!ht]
\centering
\includegraphics[width=1\linewidth]{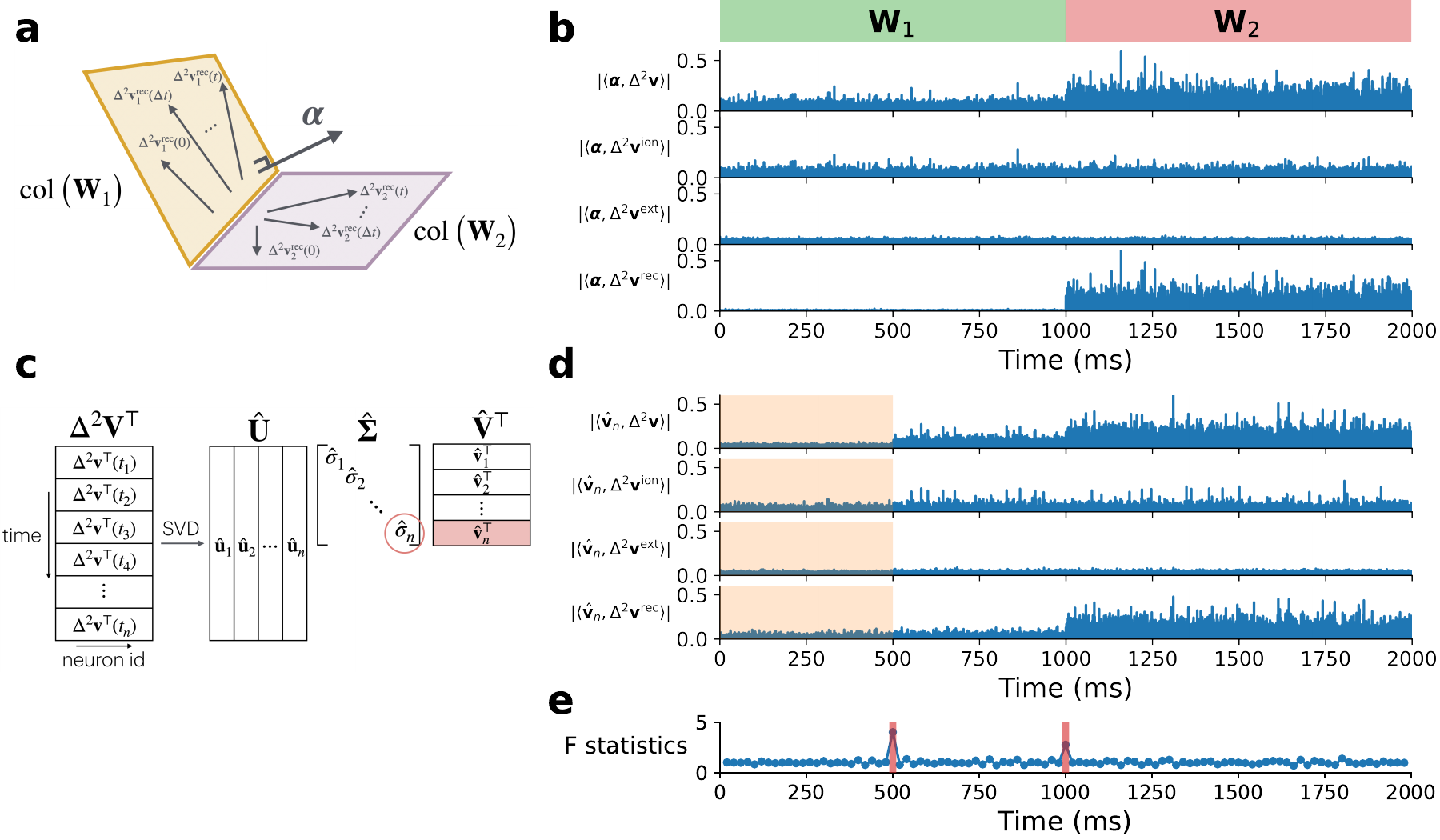}
\caption{
    Schematics of the change point detection (CPD) method.
    \textbf{a}. The schematics of subspace for $\Delta^2 \mathbf{v}^\mathrm{rec}_1$ and $\Delta^2 \mathbf{v}^\mathrm{rec}_2$.
    \textbf{b}. Time courses of projection magnitude of $\Delta^2\mathbf{v}$,
    $\Delta^2\mathbf{v}^\mathrm{ion}$,
    $\Delta^2\mathbf{v}^\mathrm{ext}$
    and $\Delta^2\mathbf{v}^\mathrm{rec}$ to the (approximated) orthogonal vector 
    ($\boldsymbol{\alpha}$) of $\mathrm{col}(\mathbf{W}_1)$, respectively (from top to bottom).
    \textbf{c}. The illustration of estimating the orthogonal vector of the subspace of spanned by 
    $\Delta^2\mathbf{v}(t)$ using singular vector decomposition.
    \textbf{d}. The projection magnitude of $\Delta^2\mathbf{v}$,
    $\Delta^2\mathbf{v}^\mathrm{ion}$,
    $\Delta^2\mathbf{v}^\mathrm{ext}$
    and $\Delta^2\mathbf{v}^\mathrm{rec}$ to last right singular vector 
    ($\hat{\mathbf{v}}$), associated with the smallest singular value, of $\Delta^2\mathbf{V}^\top$,
    with the same panel configuration as those in \textbf{b}.
    $\hat{\mathbf{v}}$ is computed using the first 500 ms $\Delta^2 \mathbf{v}(t)$, indicated by orange shaded areas in each panel.
    In \textbf{b} and \textbf{d}, the structural connectivity matrix changes from 
    $\mathbf{W}_1$ to $\mathbf{W}_2$ at 1000 ms.
    \textbf{e}. The F statistics for the projection magnitude
    $\left|\langle\hat{\mathbf{v}}_n,\Delta^2\mathbf{v}\rangle\right|$ with window size
    $20$ ms. The red shaded areas indicate the detected change points with $p<10^{-20}$. 
}
\label{fig:tcp}
\end{figure}

Consider a general EI balanced network with a change of structural connectivity matrix at time $t_c$.
This change of structural connectivity pattern might result from the growth of new axonal projections between neurons
or synaptic plasticity during learning and memory. Here, we assume the sudden change of network's structures,
considering the fact that short time synaptic plasticity happens with short time period, much less than the time duration of total recordings of neuronal activities.
For better illustration, we denote the network structural connectivity as $\mathbf{W}_1$ and $\mathbf{W}_2$ before and 
after $t_c$ (as denoted as green and red bars above the Fig. \ref{fig:tcp}b).
We rewrite Eq. \ref{eq:dv_int} in the vector form as
\begin{equation}
\begin{aligned}
    \Delta\mathbf{v}&:=\mathbf{v}(t+\Delta t)-\mathbf{v}(t) \\
    &= \frac{1}{C_m}\left\{\int_t^{t+\Delta t}
    \mathbf{I}_{\mathrm{ion}}(t')dt' 
    + \mathbf{F}\mathbf{d}^P(t)
    + \mathbf{W}\mathbf{d}(t)\right\},
\end{aligned}
\label{eq:dv_vec}
\end{equation}
where $\mathbf{F}=\mathrm{diag}(F_1, F_2, \cdots, F_N)$ and $\mathbf{W}$ is the structural connectivity matrix, taking as $\mathbf{W}_1$ before time $t_c$ and $\mathbf{W}_2$ after time $t_c$. $\mathbf{d}^P(t)$ and $\mathbf{d}(t)$ are $N$-dimensional Bernoulli 
random vectors for feedforward Poisson inputs and recurrent synaptic inputs, respectively.
Similarly, we take another temporal difference to obtain the second-order difference of the voltage vector,
defined by
\begin{equation}
\begin{aligned}
    \Delta^2\mathbf{v}&:=\mathbf{v}(t+2\Delta t)-2\mathbf{v}(t+\Delta t)+\mathbf{v}(t) \\
    &= \frac{1}{C_m}\left\{\underbrace{\int_t^{t+\Delta t}
    (\mathbf{I}_{\mathrm{ion}}(t'+\Delta t)-\mathbf{I}_{\mathrm{ion}}(t'))dt'}_{\Delta^2 \mathbf{v}^\mathrm{ion}}\right. \\
    &\left.\quad+ \underbrace{\mathbf{F}[\mathbf{d}^P(t+\Delta t) - \mathbf{d}^P(t)]}_{\Delta^2 \mathbf{v}^\mathrm{ext}}
    + \underbrace{\mathbf{W}[\mathbf{d}(t+\Delta t) - \mathbf{d}(t)]}_{\Delta^2 \mathbf{v}^\mathrm{rec}}\right\}.
\end{aligned}
\label{eq:ddv_vec}
\end{equation}
For clarity, we denote that
$$
\Delta^2\mathbf{v}(t)=\left\{
\begin{aligned}
&\Delta^2\mathbf{v}_1(t) & t\le t_c\\
&\Delta^2\mathbf{v}_2(t) & t> t_c\\
\end{aligned}\right..
$$
Notably, $\Delta^2\mathbf{v}$ depicts the population averaged fluctuation of driving currents to the network.
In the last section, we demonstrate that the variance of  $\langle\Delta^2v_i^\mathrm{rec}\rangle_i$ 
dominates the fluctuation of $\langle\Delta^2v_i\rangle_i$. Consequently, the L-2 norm of $\Delta^2\mathbf{v}$ is predominantly contributed by the 
$\Delta^2\mathbf{v}^\mathrm{rec}$.
According to Eq. \ref{eq:ddv_vec}, 
$\Delta^2 \mathbf{v}^\mathrm{rec} \in \mathrm{col}(\mathbf{W})$, resulting in
$\Delta^2\mathbf{v}^\mathrm{rec}(t)$ lying in different subspaces before and after $t_c$:
\begin{equation}
\Delta^2\mathbf{v}^\mathrm{rec}(t) = \left\{
\begin{aligned}
&\Delta^2\mathbf{v}_1^\mathrm{rec}(t)\in\mathrm{col}(\mathbf{W}_1) &t\le t_c\\
&\Delta^2\mathbf{v}_2^\mathrm{rec}(t)\in\mathrm{col}(\mathbf{W}_2) &t> t_c\\
\end{aligned}
\right..
\end{equation}
In $N$-dimensional space, $\mathrm{col}(\mathbf{W}_1)$ and $\mathrm{col}(\mathbf{W}_2)$ are not overlap with each other in high probability, as illustrated in Fig. \ref{fig:tcp}\textbf{a}.
Assuming that $\boldsymbol{\alpha}$ is a orthogonal vector with respect to $\mathrm{col}(\mathbf{W}_1)$, the projection magnitude of $\Delta^2\mathbf{v}_1^\mathrm{rec}$ onto $\boldsymbol{\alpha}$, \textit{i.e.} $|\langle\boldsymbol{\alpha}, \Delta^2\mathbf{v}_1^\mathrm{rec}\rangle|$,
approaches to zero, while $|\langle\boldsymbol{\alpha}, \Delta^2\mathbf{v}_2^\mathrm{rec}\rangle|$ remains significantly nonzero, as shown in Fig.
\ref{fig:tcp}\textbf{b}(bottom). Likewise, the difference of projections onto $\boldsymbol{\alpha}$
between $\Delta^2\mathbf{v}_1$ and $\Delta^2\mathbf{v}_2$ is also prominent, 
as shown at the top of Fig. \ref{fig:tcp}\textbf{b}.
This apparent mismatch of projections act as an indicator for the difference 
of connectivity, helping us to identify the time point of the connectivity change.
For the case in Fig. \ref{fig:tcp}\textbf{b}, the change occurs at time 1000 ms.
In addition, the $\Delta^2\mathbf{v}^\mathrm{ion}$ and $\Delta^2\mathbf{v}^\mathrm{ext}$ are 
weakly correlated with $\Delta^2\mathbf{v}^\mathrm{rec}$, resulting in less prominent change in
their projection magnitudes onto $\boldsymbol{\alpha}$ before and after change point,
as shown in Fig. \ref{fig:tcp}\textbf{b} 
(two middle panels).

However, in practical applications, we have no access to $\mathbf{W}_1$. To address this, 
we approximate $\mathrm{col}(\mathbf{W}_1)$ with the subspace spanned by $\{\Delta^2\mathbf{v}(t)\}$.
Consequently, in our CPD algorithm, we proxy $\boldsymbol{\alpha}$ with the 
right singular vector $\hat{\mathbf{v}}_n$ of $\Delta^2\mathbf{V}^\top$ associated with the smallest singular value,
as illustrated in Fig. \ref{fig:tcp}\textbf{c}.
In Fig. \ref{fig:tcp}\textbf{d}, we demonstrate the performance of $\hat{\mathbf{v}}_n$ estimated
using the first $500$ ms voltage time series. 
The projection magnitudes onto $\hat{\mathbf{v}}_n$ works similarly as those onto
$\boldsymbol{\alpha}$. The change point at time $1000\,\mathrm{ms}$ can still be clearly detected using the metric
$|\langle\hat{\mathbf{v}}_n,\Delta^2\mathbf{v}\rangle|$.

To quantitatively identify exact time of the change point, we perform the following statistical test on the projection of $\Delta^2\mathbf{v}$. Since we've proved the
normal-distributed feature of $\left|\langle\Delta^2v_i\rangle_i\right|$ at large-$N$ limit,
the distribution of projection of $\Delta^2\mathbf{v}$ onto a given vector $\hat{\mathbf{v}}_n$ also 
converges to normal distribution according the central limit theorem. 
From Fig. \ref{fig:tcp}d, we notice that the variance of $\langle\hat{\mathbf{v}}_n, \Delta^2\mathbf{v}\rangle$ shifts significantly before and after the change point. 
Based on this observation, we segment the time series of 
$\langle\hat{\mathbf{v}}_n, \Delta^2\mathbf{v}\rangle(t)$ into non-overlapping equal-length pieces
using a time window $\Delta T$. 
We then test the null hypothesis that the sample variances of two consecutive segments are
equal, using the following \textit{F}-statistic: 
\begin{equation}
\begin{aligned}
    F(t)&=\frac{
    \frac{1}{\Delta T/\Delta t-1}
    \sum\limits_{t'=t}^{t+\Delta T/\Delta t}
    \left[\langle\hat{\mathbf{v}}_n, \Delta^2\mathbf{v}\rangle(t')
    -\mathbb{E}_{t''\in[t,t+\Delta T/\Delta t]}\left[\langle\hat{\mathbf{v}}_n, \Delta^2\mathbf{v}\rangle(t'')\right]\right]^2}
    {\frac{1}{\Delta T/\Delta t-1}
    \sum\limits_{t'=t-\Delta T/\Delta t}^{t}
    \left[\langle\hat{\mathbf{v}}_n, \Delta^2\mathbf{v}\rangle(t')-
    \mathbb{E}_{t''\in[t-\Delta T/\Delta t,t]}\left[\langle\hat{\mathbf{v}}_n, \Delta^2\mathbf{v}\rangle(t'')\right]\right]^2}, \\
    t&=\left\{\Delta T, 2\Delta T, \cdots, 
    \left(\mathrm{round}\left(\frac{T}{\Delta T}\right)-1\right)\cdot\Delta T\right\},
\end{aligned}
\end{equation}
where $\Delta t$ is the size of sampling time step and $T$ is the recording duration.
This \textit{F}-statistic follows the \textit{F}-distribution with degrees of freedom 
$d_1=d_2=\Delta T/\Delta t-1$ under the null hypothesis.
The null hypothesis is rejected when a change point occurs at time $t_c$,
resulting in $F(t_c)$ exceeding the critical \textit{F} value.
As shown in Fig. \ref{fig:tcp}\textbf{e}, \textit{F} statistics exhibits a pronounced peak
at time $1000$ ms, highlighted by the red shaded area, accurately identifying the change point.
We note that an additional peak appears at $500$ ms, which stems from using the first $500$ ms voltage times series to estimate the projection vector.
This false alarm can be eliminated by adjusting the data length used for analysis, demonstrating the robustness of our change point detection method. 

\subsection{Network reconstruction using time-delayed correlation-coefficient}\label{sec4}

\begin{figure}[!h]
\centering
\includegraphics[width=1\linewidth]{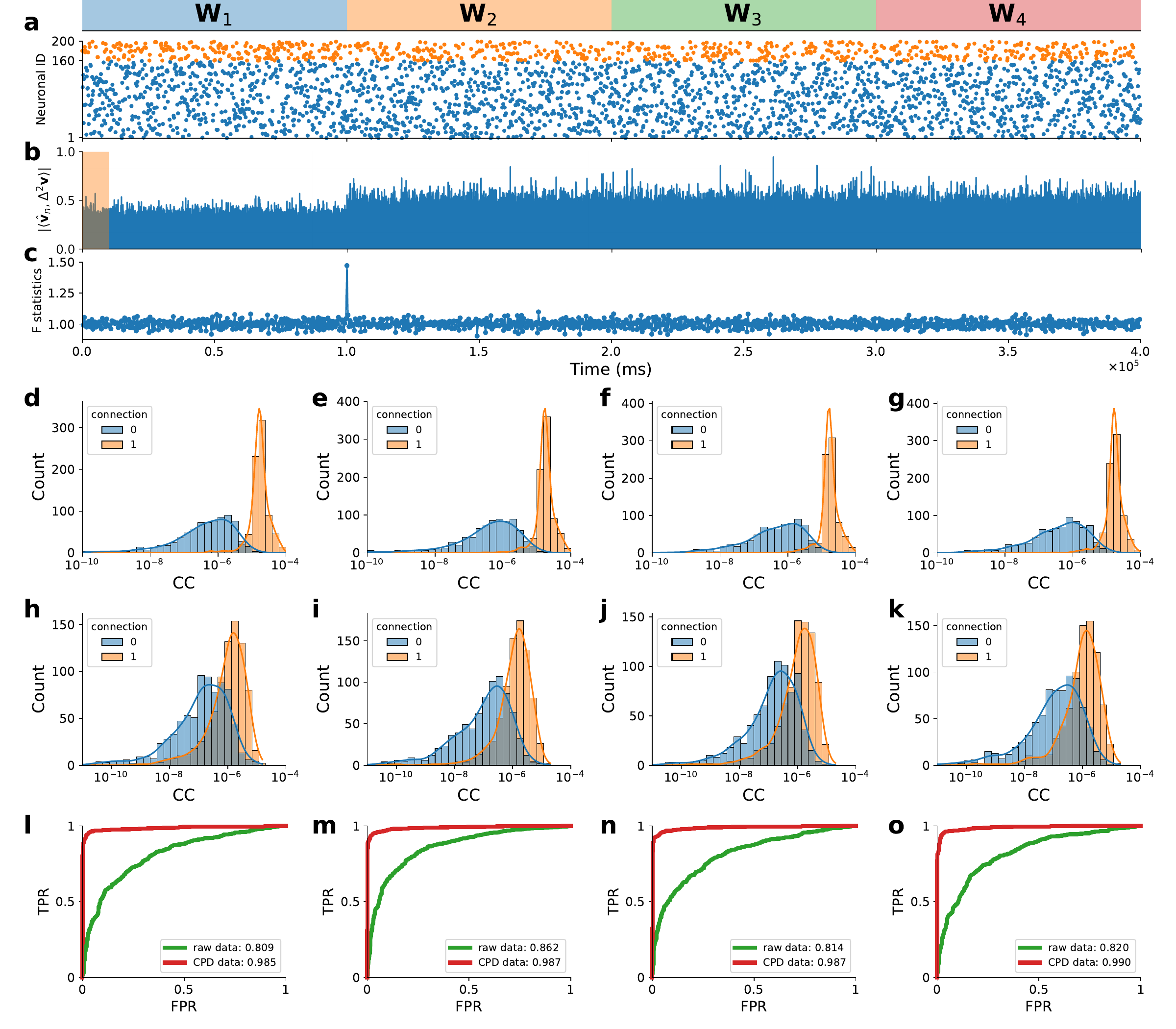}
\caption{
    CPD-TDCC network reconstruction for LIF neuronal network in the balanced regime.
    \textbf{a}. Raster plot for an 4000-neuron LIF network with three connectivity-induced change points.
    Only 0.1 \% spikes are plotted for the easy of visualization.
    Top: the colored bar indicates 
    the timeline of the network connectivity matrix transitions, ranging from 
    $\mathbf{W}_1$ to $\mathbf{W}_4$.
    \textbf{b}. Time course of inner product magnitude between $\Delta^2 \mathbf{v}(t)$ and $\hat{\mathbf{v}}$.
    The projection vector $\hat{\mathbf{v}}$ is computed using the first 10 s of 
    $\Delta^2 \mathbf{v}(t)$, indicated by orange shaded areas.
    \textbf{c}. The F statistics for the projection
    $\langle\hat{\mathbf{v}}_n,\Delta^2\mathbf{v}\rangle$ in \textbf{b} with window size
    $400$ ms. The pronounced peak at time $10^{5}$ ms indicates the detected change point,
    with $p<10^{-20}$.
    \textbf{d}-\textbf{g}. Performance of network reconstructions using CPD-TDCC method.
    The histograms of pairwise TDCC of 4\% of random subsets of neuron pairs in the 200-neuron
    subnetwork,
    using connectivity matrix
    (\textbf{d}) $\mathbf{W}_1$,  
    (\textbf{e}) $\mathbf{W}_2$,
    (\textbf{f}) $\mathbf{W}_3$ and
    (\textbf{g}) $\mathbf{W}_4$ as ground truth, respectively.
    In these result, TDCC values are estimated using $1\times10^5$ ms segmented spike trains
    between successive change points.
    \textbf{h}-\textbf{k}. Performance of network reconstructions using native TDCC method,
    organized similarly as those in \textbf{c}-\textbf{f}.
    TDCC values are estimated using total $4\times10^5$ ms spike trains for all four results.
    \textbf{l}-\textbf{o}. The ROC curves quantifying the performance of the TDCC-based
    network reconstruction using native TDCC (green curves) and CPD-TDCC method (red curves).
    The AUC values for each case are included in legends.
    The parameters of TDCC are chosen as $\Delta t=0.1\,\mathrm{ms}$, $\tau=0.1\,\mathrm{ms}$.
}
\label{fig:recon}
\end{figure}

In this section, we demonstrate that the effectiveness of our CPD-TDCC network reconstruction
framework and method and emphasize the performance improvement by introducing
CPD method.
We simulate a 4000-neuron LIF neuronal network, comprising of 3200 excitatory and 800 inhibitory
neurons for $4\times 10^5\,\mathrm{ms}$ with time step $\Delta t=0.02\,\mathrm{ms}$.
The network parameters for coupling strength and Poisson input are given as those in Table \ref{tab1}.
During simulation, the connectivity matrix $\mathbf{W}$ of the network
changes every $10^5\,\mathrm{ms}$, denoted as $\mathbf{W}_1$ to $\mathbf{W}_4$,
as shown in the top of Fig. \ref{fig:recon}\textbf{a},
producing three change points at time $1\times10^5\,\mathrm{ms}$, $2\times10^5\,\mathrm{ms}$, and
$3\times10^5\,\mathrm{ms}$, respectively.
Considering the fact that neurons in cortical network can only be sparsely labeled and recorded,
we sample a 200-neuron subnetwork within the large network randomly, and record voltage
time series and spike trains of those neurons.
The network's mean firing rate is around $52\,\mathrm{Hz}$, and the raster of recorded neurons
are shown
in Fig. \ref{fig:recon}\textbf{a} (only 0.1\% randomly sampled spikes are plotted for the ease of illustration).
Following our CPD framework, we use the first 10-second voltage time series to estimate the projection vector
$\hat{\mathbf{v}}_n$, as indicated by orange shaded areas in Fig. \ref{fig:recon}\textbf{b}. The 
projection magnitude of $\Delta^2\mathbf{v}$ onto $\hat{\mathbf{v}}_n$ exhibits a sharp transition
of mean values at $t=100\,\mathrm{s}$, as shown in Fig. \ref{fig:recon}\textbf{b}.
Quantitatively, we perform the F-test to accurately identify the first change point 
with time course of F statistics illustrated in Fig. \ref{fig:recon}\textbf{c}, showing
a delta peak at $t=100$ s.
We use the binarized spike-train time series of the first 100 seconds to estimate TDCC values of
all neuronal pairs in the subnetwork. The histogram of TDCC values for connected (orange) and 
unconnected (blue, only 2\% of randomly sampled pairs included in analysis) pairs are almost separated, as shown in Fig. \ref{fig:recon}\textbf{d},
indicating a good reconstruction performance. The ROC analysis gives a 0.985 AUC values,
as shown by the red curve in Fig. \ref{fig:recon}\textbf{l}.
To further quantify the performance improvement attributed from CPD, we also perform the
TDCC-based network reconstruction using spike trains of all 400 seconds. As shown in Fig.
\ref{fig:recon}\textbf{h}, the histogram of connected and unconnected pairs largely 
overlap, leading to a worse reconstruction performance with a 0.809 AUC value (green curve
in Fig. \ref{fig:recon}\textbf{l}).
As a result, CPD-TDCC reconstruction framework achieves a 0.176 AUC improvement compared
with native TDCC-based reconstruction, emphasizing the effectiveness of our CPD-TDCC reconstruction
framework.

Our reconstruction pipeline can be recursively applied for identifying the 
resting change points and reconstructing remaining three connectivity matrix ($\mathbf{W}_2$-$\mathbf{W}_4$).
Using the voltage time series within time period $t\in$[100, 110] seconds, our CPD method can identify the second
change point at $t=200$ seconds. We compute the TDCC using binarized spike trains between 
the first and second change points, \textit{i.e.}, from 100 to 200 seconds,
and well reconstruct the structural connectivity (see Fig. \ref{fig:recon}\textbf{e)},
achieving 0.987 AUC value (red curve in Fig. \ref{fig:recon}\textbf{m}).
As a comparison, TDCC-based reconstruction without CPD only achieves a 0.862 AUC value 
(see Fig. \ref{fig:recon}\textbf{i} and green curve in \textbf{m}).
Similarly, performances of reconstructing $\mathbf{W}_3$ and $\mathbf{W}_4$ are shown in Fig.
\ref{fig:recon}\textbf{f}-\textbf{g} using CPD-TDCC pipeline, respectively, which is
significantly better than those of native TDCC-based reconstructions (see in Fig. \ref{fig:recon}\textbf{j}-\textbf{k}), as shown in Fig. \ref{fig:recon}\textbf{n}-\textbf{o}.

\subsection{Reconstruction of general balanced neuronal networks}\label{sec:HH}

\begin{figure}[!h]
\centering
\includegraphics[width=1\linewidth]{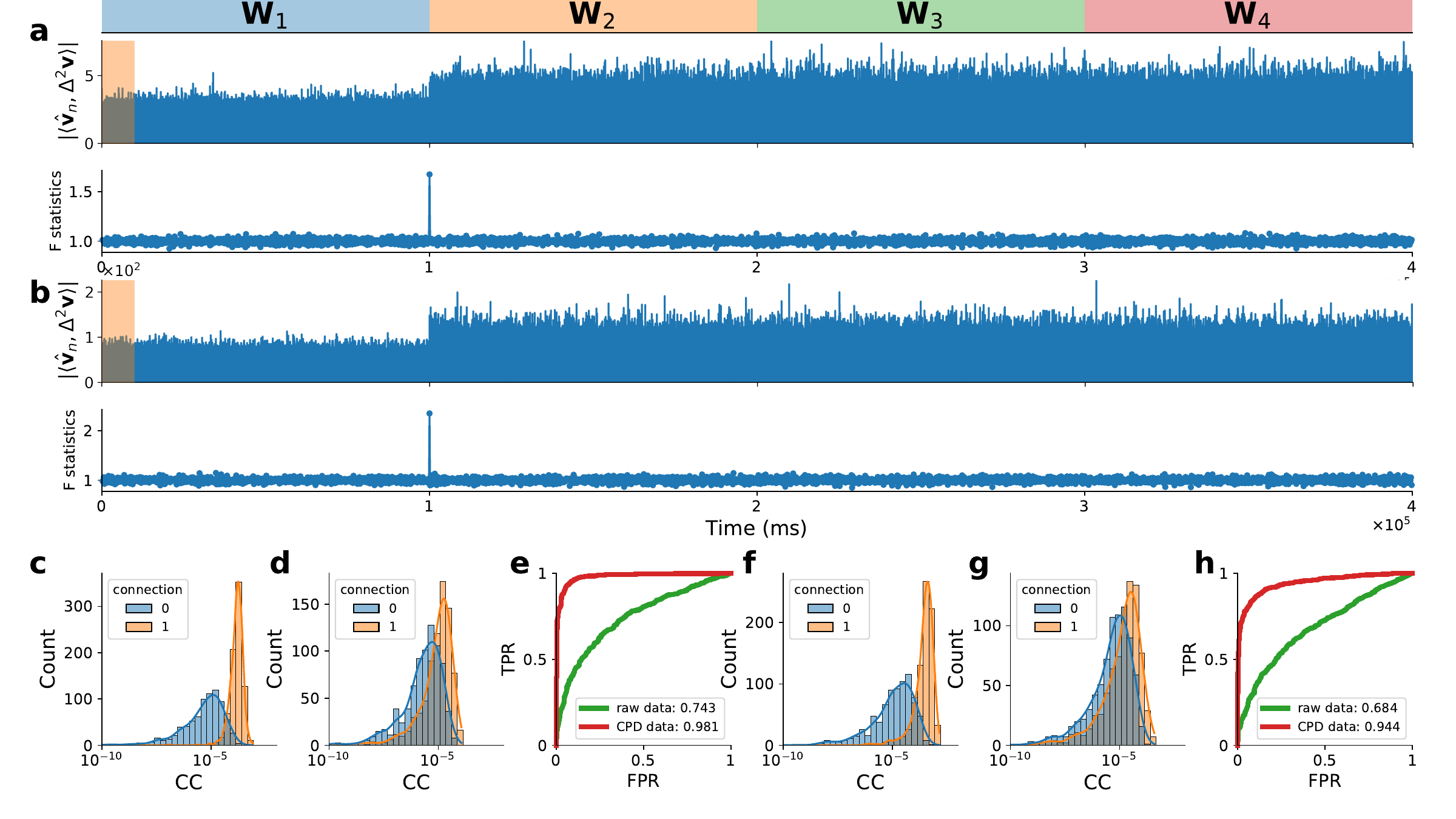}
\caption{
    CPD-TDCC network reconstruction for 4000-neuron Hodgkin-Huxley network and Morris-Lecar network in balanced regimes.
    \textbf{a}. (Top) Time course of inner product magnitude between $\Delta^2 \mathbf{v}(t)$ and $\hat{\mathbf{v}}$, in the HH network.
    The projection vector $\hat{\mathbf{v}}$ is computed using the first 10 s of 
    $\Delta^2 \mathbf{v}(t)$, indicated by orange shaded areas.
    (Bottom) The F statistics for the projection in \textbf{a} with window size
    $200$ ms. The pronounced peak at $10^{5}$ ms with $p<10^{-20}$ identifies the change point.
    \textbf{b}. Time course of projection magnitudes and F statistics, same as those
    in \textbf{a}, but for the Morris-Lecar network.
    \textbf{c}-\textbf{e}. Network reconstruction results for Hodgkin-Huxley network.
    \textbf{c}. Histogram of pairwise TDCC among connected (orange) and 
    unconnected (blue) neuronal pairs, respectively,
    estimated using spike train 
    recordings before the first change point, \textit{i.e.}, $t\in[0,100\,\mathrm{s}]$. 
    \textbf{d}. Histogram of pairwise TDCC among connected (orange) and 
    unconnected (blue) neuronal pairs, respectively,
    but estimated using spike train 
    recordings using total $400\,\mathrm{s}$ spike trains. 
    \textbf{e}. The ROC curves of using TDCC as the metric for reconstructing network 
    adjacency matrix, based on results in \textbf{b} (red) and 
    \textbf{c} (green), respectively.
    \textbf{f}-\textbf{h}. Network reconstruction results organized the same as those in
    \textbf{c}-\textbf{e}, but for Morris-Lecar network.
    Parameters of TDCC estimation are chosen as $\Delta t=0.5\,\mathrm{ms}$, $\tau=0.1\,\mathrm{ms}$.
}
\label{fig:recon_general}
\end{figure}

To further verify the applicability of our CPD-TDCC reconstruction pipeline on real neural
systems, we apply it to reconstruct Hodgkin-Huxley (HH) neuronal networks in balanced regime.
We simulate a 4000-neuron HH network consisting of 3200 excitatory and 800 inhibitory neurons.
Similar as the LIF network in Fig. \ref{fig:recon},
the HH network also has three change points, as shown in the top of Fig. \ref{fig:recon_general}\textbf{a}, denoting as $\mathbf{W}_1$-$\mathbf{W}_4$. 
Those change points also occur every $1\times 10^5$ ms, altering the causal relations between neurons.
Similarly, we use the first 10 second voltage time series to estimate the projection vector 
$\hat{\mathbf{v}}_n$, and the mean projection magnitude of $\Delta^2\mathbf{v}$ shows a sharp
transition at $t=100$ seconds (see middle panel in Fig. \ref{fig:recon_general}\textbf{a}), locating the change point (identified by \textit{F}-test with $p<10^{-20}$, bottom of Fig. \ref{fig:recon_general}\textbf{a}). 
With the identified change point, we compute the pairwise TDCC values using the first $1\times10^5$ ms binarized spike trains for reconstructing the connectivity matrix $\mathbf{W}_1$.
As shown in Fig. \ref{fig:recon_general}\textbf{c}, the histogram of pairwise TDCC values for 
connected and unconnected pairs almost separate apart, achieving 0.981 AUC value (see Fig. \ref{fig:recon_general}\textbf{e}).
Compared with native TDCC-based reconstruction using spike trains of all 400 seconds data as 
shown in Fig. \ref{fig:recon_general}\textbf{d}, our CPD-TDCC reconstruction pipeline again shows
significant advantages in dealing with network reconstruction with connectivity-induced change points.
We also test its performance on a 4000-neuron Morris-Lecar neuronal network in the balanced regime.
As shown in Fig. \ref{fig:recon_general}\textbf{b} (top), the change point at 100
seconds are accurately captured using $\hat{\mathbf{v}}_n$ estimated from the first 10-second
voltage recordings. Consistently, the reconstruction performance for $\mathbf{W}_1$ of using 
CPD-TDCC pipeline (see Fig.~\ref{fig:recon_general}\textbf{f}) significantly outperforms those of
using native TDCC-based reconstructions (see Fig.~\ref{fig:recon_general}\textbf{g}), as shown by the
increase of AUC value by 0.26 in Fig. \ref{fig:recon_general}\textbf{h}.

\subsection{Reconstruction networks with weight changes}

Previous successful applications of our CPD-TDCC reconstruction framework have focused on 
scenarios where network adjacency matrices change at change points,
while the coupling strength between connected neuronal pairs remains fixed.
This approach effectively models the growth of axonal
projections in cortical networks, a process that occurs less frequently in adult brains.
Meanwhile, the synaptic coupling strength undergoes continuous modification due to
the synaptic plasticity during learning and memory formation. Therefore, extending our methods
to detect changes in coupling strength, rather than merely connectivity structure, represents
a critical advancement for understanding the network structure in mature brains.  

\begin{figure}[!ht]
    \centering
    \includegraphics[width=1\linewidth]{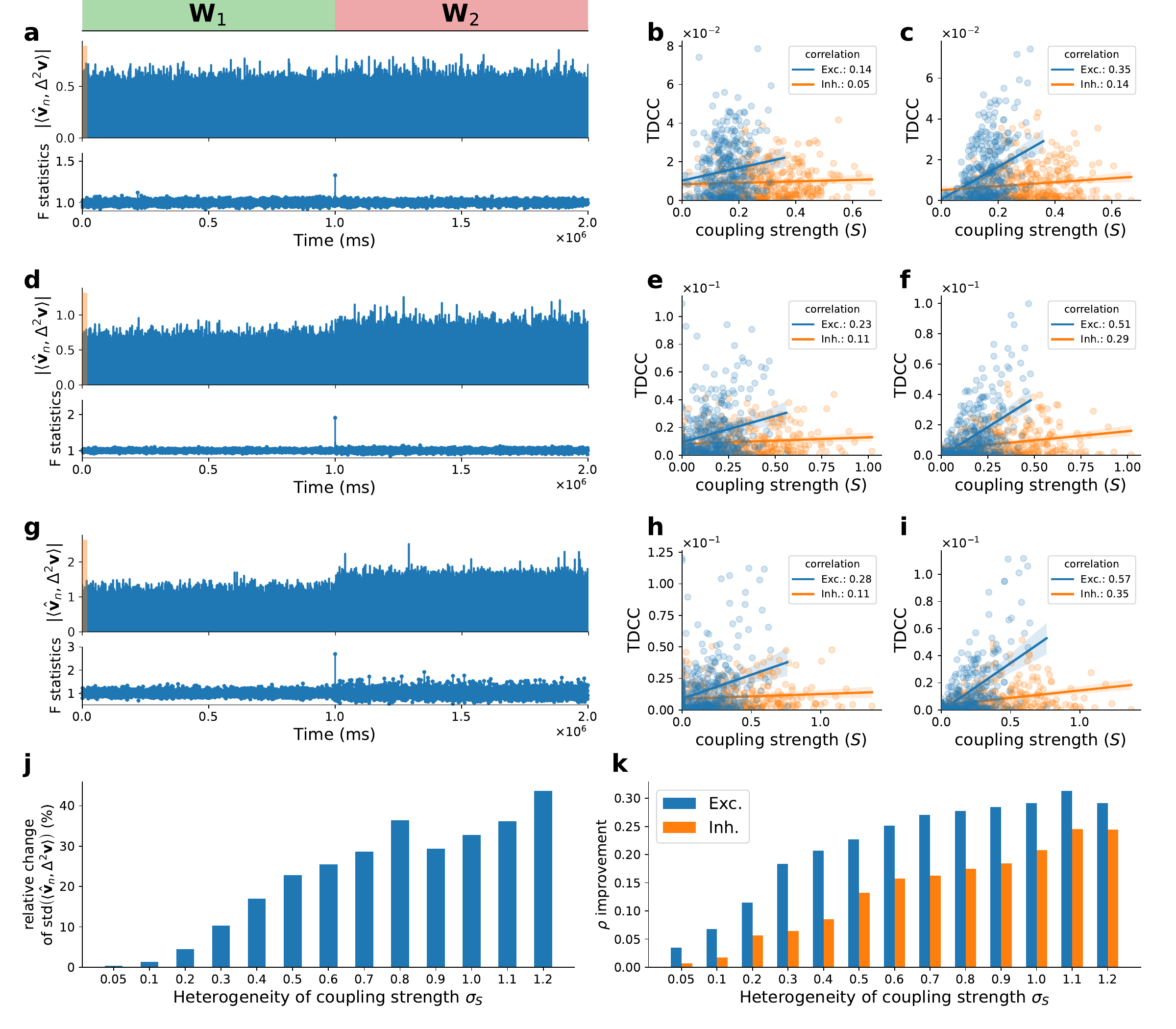}
    \caption{
    CPD-TDCC network reconstruction for LIF networks with heterogeneous coupling strength parametrized with variance $\sigma_S$. \textbf{a}.
    (Top) Time course of inner product magnitude between $\Delta^2 \mathbf{v}(t)$ and $\hat{\mathbf{v}}$ for the network with $\sigma_S=0.4$.
    The projection vector $\hat{\mathbf{v}}$ is computed using the first 10 s of 
    $\Delta^2 \mathbf{v}(t)$, indicated by orange shaded areas.
    (Bottom) The corresponding F statistics for the projection magnitude
    with window size $500$ ms.
    \textbf{b}. Relationships between coupling strength $S$ and pairwise TDCC value
    for connected pairs among excitatory (blue) and inhibitory (orange) populations, respectively, estimated using the total 4000 s spike train recordings.
    The corresponding colored lines are linear regressions with shaded areas indicating
    95\% confidence intervals.
    The correlation coefficients between $S$ and TDCC values are denoted in legends.
    \textbf{c}. The same results as in \textbf{b} but with TDCC values estimated using
    recordings before the change point, \textit{i.e.}, $t\in[0,1000\,\mathrm{s}]$. 
    \textbf{d-f}. Results organized similarly as those in \textbf{a}-\textbf{c} but for $\sigma_S=0.8$.
    \textbf{g-i}. Results organized similarly as those in \textbf{a}-\textbf{c} but for $\sigma_S=1.2$.
    In \textbf{d} (bottom) and \textbf{g} (bottom), the pronounced peaks at $10^{6}$ ms
    with $p<10^{-20}$ identify the change points.
    \textbf{j}. The relative variance change of
    $\langle\hat{\mathbf{v}}_n, \Delta^2 \mathbf{v}\rangle(t)$ before and after the change point
    for LIF networks with different strength heterogeneity $\sigma_S$.
    \textbf{k}.
    The improvement of correlation coefficient $\rho$ between coupling strength and TDCC values,
    for LIF networks with different strength heterogeneity $\sigma_S$.
    Parameters of TDCC estimations are chosen as $\Delta t=0.5\,\mathrm{ms}$, $\tau=0.5\,\mathrm{ms}$.
    }
    \label{fig:lif_vary_weight}
\end{figure}

Here, we demonstrate the effectiveness of our framework applying on balanced-state networks with 
changing synaptic coupling strength while maintaining fixed network topology (adjacency matrix).
We again simulate 4000-neuron LIF networks for $2\times 10^6\,\mathrm{ms}$, with connectivity
matrices transitioning from $\mathbf{W}_1$ to $\mathbf{W}_2$ at $t=10^6\,\mathrm{ms}$,
as illustrated in Fig. \ref{fig:lif_vary_weight}\textbf{a} (top). In this case, we have
\begin{equation}
\mathbf{W}_1=\mathbf{S}_1\odot\mathbf{A}\quad\mathrm{and}\quad
\mathbf{W}_2=\mathbf{S}_2\odot\mathbf{A},
\end{equation}
where $\mathbf{A}$ represents the unchanged network adjacency matrix, and $\odot$ represents the Hadamard product.
$\mathbf{S}_1$ and $\mathbf{S}_2$ are weight matrices, sampled from identical sets
of independent normal distributions before and after the change point, respectively.
For each weight matrix $\mathbf{S}_k$ ($k=\{1,2\}$), 
the coupling weight from neuron $i$ of type $Q_i$ to neuron
$j$ of type $Q_j$ is independently sampled from a normal distribution
$S_{k, ij}^{Q_iQ_j}\sim\mathcal{N}(J^{Q_iQ_j}, (J^{Q_iQ_j}\sigma_S)^2)$, where $\sigma_S$ controls
the heterogeneity of coupling weights.
Larger $\sigma_S$ values produce greater the temporal change cross the change point.
The neuron type-dependent weight factors $J^{Q_iQ_j}$ are consistent with those 
specified in Table \ref{tab1} for LIF networks.

According to our theoretical framework for CPD in Sec. \ref{sec:CPD}, the discrepancy between
$\mathbf{S}_1$ and $\mathbf{S}_2$ induces a misalignment between the subspaces 
$\mathrm{col}(\mathbf{W}_1)$ and $\mathrm{col}(\mathbf{W}_2)$,
thereby validating the effectiveness of the CPD framework.
Intuitively, larger changes in coupling strength (characterized by higher values of 
heterogeneity $\sigma_S$)
leads to greater subspace misalignment.
To investigate this relationship, we systematically vary the coupling strength 
heterogeneity $\sigma_S$ from $0.05$ to $1.2$, and then use first 10 seconds of 
voltage time series to estimate the projection vectors for CPD analysis.
For relatively small heterogeneity ($\sigma_S=0.4$),
as shown in Fig. \ref{fig:lif_vary_weight}\textbf{a}, the average projection magnitudes
before and after the change point remain virtually indistinguishable. Nevertheless,
the \textit{F}-test still successfully identifies the change point at $10^6$ ms.
As the level of change increases, shown in Fig. \ref{fig:lif_vary_weight}\textbf{d}
($\sigma_S=0.8$) and \textbf{g} ($\sigma_S=1.2$), change point
at $10^6\,\mathrm{ms}$ becomes increasingly evident and is robustly detected by the corresponding \textit{F}-test.
To quantify the detectability of our CPD across different level of structural changes,
we compute the standard deviation of projection
$\langle \hat{\mathbf{v}}_n, \Delta^2\mathrm{v}\rangle$ separately for the first half
($t\in[0,10^6)\,\mathrm{ms}$) and the second half ($t\in[10^6, 
2\times10^6]\,\mathrm{ms}$) of the voltage recordings,
across different $\sigma_S$ values.
As demonstrated in Fig. \ref{fig:lif_vary_weight}\textbf{j}, the relative difference of the standard deviation increases monotonically with the level of coupling strength
changes.
For small variations in coupling strength, specifically when $\sigma_S < 0.2$,
the changes are too subtle to be reliably detected by our CPD framework, addressing the feasible range of our method.

We further quantify how the CPD enhances the performance of network reconstruction. Since the network
adjacency matrix remains unchanged across the structural change point, ROC-AUC analysis,
designed for binary classification, is not appropriate in this context.
Instead, inspired by the linear correlation between
TDCC values and coupling strengths established in Sec. \ref{sec:recon_theory},
we adapt this correlation as our performance metric.
Specifically, we first compute TDCC of all connected neuronal pairs within a 200-neuron subnetwork,
using voltage time series and spike trains recorded over the entire $2\times 10^6\,\mathrm{ms}$ 
simulation. We then compute the Pearson correlation coefficient $\rho$ between TDCC values and their
underlying coupling weight from $\mathbf{S}_1$. As shown in Fig. \ref{fig:lif_vary_weight}\textbf{b},
TDCC exhibits show weak correlations with coupling strength for
both excitatory ($\rho=0.14$, blue) and inhibitory ($\rho=0.05$, orange) connections.
In contrast, with the help of CPD method, the correlation between TDCC values and $\mathbf{S}_1$ 
improves substantially for both excitatory connections ($\rho=0.35$) and inhibitory connections ($\rho=0.14$), as shown in Fig. \ref{fig:lif_vary_weight}\textbf{c}.
This indicates that isolating data segments through CPD before the structural change yields more accurate reconstruction.
For networks with larger coupling weight changes, CPD-TDCC method makes even greater improvements
in those correlations, as demonstrated in Fig. \ref{fig:lif_vary_weight}\textbf{e},
\textbf{f}, \textbf{h} and \textbf{i}. 
Overall, we observe that the correlation improvement scales positively correlated with
the degree of coupling heterogeneity,
as shown in Fig. \ref{fig:lif_vary_weight}\textbf{k},
suggesting that CPD yields greater benefits in scenarios involving more substantial structural changes.

\section{Discussion}\label{sec:discussion}

We introduce a connectivity-induced change point detection and time-delayed correlation 
coefficient framework for reconstructing balanced neuronal networks with structural connectivity modifications.
First, we developed a theoretical framework for detecting connectivity-induced change points based on the scaling properties of balanced-state networks and the resulting fluctuations in membrane potentials.
The order of magnitude and the geometric properties the fluctuations of membrane potentials
help identify changes of neuronal dynamics induced by modifications of underlying structural connectivity patterns. 
Second, we demonstrated that segmenting neural recordings at these detected change points prior to
applying TDCC-based reconstruction significantly improves performance compared to
conventional TDCC methods that assume static structural connectivity.
Our simulations with multiple types of neuronal network models in balanced regimes revealed that the CPD-TDCC 
framework effectively reconstructs both changes in network adjacency matrices and modulations
in synaptic coupling strengths. 
Notably, changes in coupling strengths within the fixed underlying network topology closely mimics
synaptic plasticity happens in mature brains,
and the ability to detect and account for such scenarios suggests that our method is particularly
valuable for analyzing complex, heterogeneous neural circuits.

Previous change point detection methods have also explored the possibility of detecting change points induced by structural change
of complex systems \cite{hou2022harvesting}. However, this state space embedding method requires system smoothness,
making them inapplicable to spiking neuronal networks.
Our CPD approach addresses this critical gap by explicitly accounting for temporal change points in network structure, 
due to synaptic plasticity, for excitation-inhibition balanced spiking neuronal network, without requiring smoothness assumptions.
Our method offers several distinct advantages.
First, it utilizes the feature of the dominant recurrent-interaction-induced voltage fluctuations in balanced networks,
requiring no additional assumptions.
Our approach relies solely on the second-order difference of voltage time series to effectively detect changes of
$\Delta^2\mathbf{v}$ subspaces, induced by alternations in underlying structural connectivity,
making it both mathematically elegant and computationally efficient.
Secondly, our method can naturally generalize to networks beyond the classical balanced regime, such as inhibitory-dominant
networks. According to our theoretical analysis, as long as the scaling relation of network coupling strength and external
Poisson inputs as a function of average input degree follows the same relation as in classical balanced setups,
our CPD method remains effective.
Thirdly, as demonstrated in the numerical examples, the method can effectively detect structural connectivity changes based on
voltage recordings from a subnetwork, which requiring global information from the full network.
Combined with the effective subnetwork TDCC-based reconstruction, our CPD-TDCC framework accurately reconstructs subnetworks
without access to information of unmeasured neurons.
Given the experimental limitations of sparse labeling and recording in neural data acquisition, our methodology is
particularly suitable for network reconstruction from real experimental recordings.

Despite its demonstrated effectiveness, CPD-TDCC framework has several limitations that warrant further investigations.
First, our method may be compromised in networks driven by extremely fluctuating external inputs. According to Eq. \ref{eq:ddv}, 
the predominant contribution of voltage fluctuations from by recurrent interactions breaks down when $F_i$ reaches
$\mathcal{O}(\sqrt{K})$, potentially resulting in failure of change point detection. Nevertheless, the scaling of balanced network
are ubiquitous in cortical network, ensuring a wide applicability of our method under physiologically relevant conditions.
Second, the current implementation assumes discrete change points rather than continuous evolution of structural connectivity. 
This approach requires that the timescale of changes is substantially shorter than the duration of neural recordings.
While this approximation is reasonable for many experimental paradigms, developing methods for tracking gradually evolving connectivity remains an important direction for future work.
Third, our approach relies on access to membrane potential recordings of subnetworks to be reconstructed, which may not always be available in experimental settings. Extending the framework to work with more commonly available data types, such as calcium imaging, would broaden its applicability.

In conclusion, the CPD-TDCC framework represents a significant advancement in 
detecting connectivity-induced change points and reconstructing time-varying structural connectivity,
providing profound implications for understanding neural circuit dynamics.
Moreover, it could potentially contribute to identifying pathological changes in network structure induced by various  
neurological and psychiatric disorders, thereby advancing our understanding of disease mechanisms.
Overall, by explicitly accounting for connectivity-induced change points, this approach provides more accurate reconstructions of neuronal networks and offers new insights into the structure and dynamics of complex neuronal networks.

\backmatter


\bmhead{Acknowledgments}

This work was supported by National Key R\&D Program of China 2023YFF1204200 (S. Li, D. Zhou);
Shanghai Municipal Commission of Science and Technology with Grant No. 24JS2810400 (S. Li, D. Zhou);
National Natural Science Foundation of China Grant 12271361, 12250710674 (S. Li);
National Natural Science Foundation of China with Grant No. 12225109 (D. Zhou)
and the Student Innovation Center at Shanghai Jiao Tong University
(K. Chen, M. Wang, S. Li, D. Zhou)

\section*{Declarations}


\begin{itemize}
\item Conflict of interest: The authors have no relevant financial or non-financial interests to disclose.
\item Data and code availability: Data and code will be made available based on reasonable requests.
\end{itemize}


\begin{appendices}





\end{appendices}


\bibliography{C-TCP_and_reconstruction}

\end{document}